\documentclass[journal]{IEEEtran}

\usepackage{booktabs} \usepackage[svgnames,dvipsnames,x11names]{xcolor}
\usepackage{multirow}
\usepackage{subcaption}

\usepackage{latexsym}
\usepackage{amssymb}
\usepackage{amsfonts}
\usepackage{amsmath}
\usepackage{balance}
\usepackage{graphicx}
\usepackage{pifont}
\usepackage{paralist}
\usepackage{relsize}

\renewcommand{\t}{\texttt}

\newcommand{\para}[1]{\smallskip\noindent\textbf{#1}}

\DeclareMathAlphabet{\mathcal}{OMS}{cmsy}{m}{n}
\newcommand{\legacy}{{\sf\smaller Legacy}}
\newcommand{\simple}{{\sf\smaller Simple}}
\newcommand{\greedy}{{\sf\smaller Greedy}}
\newcommand{\optimal}{{\sf\smaller Optimal}}
\newcommand{\bigno}{{\sf\smaller 9node-450}}
\newcommand{\bigyes}{{\sf\smaller 9node-600}}
\newcommand{\medno}{{\sf\smaller 6node-450}}
\newcommand{\medyes}{{\sf\smaller 6node-600}}
\newcommand{\smallno}{{\sf\smaller 4node-450}}
\newcommand{\smallyes}{{\sf\smaller 4node-600}}

\newcommand{\ipnodeset}{\ensuremath{\mathcal{I}}}
\newcommand{\ipnodename}[1]{\ensuremath{\mathcal{I}}#1}
\newcommand{\routername}[1]{I#1}
\newcommand{\optname}[1]{O#1}
\newcommand{\edgename}[1]{E#1}
\newcommand{\routerset}{\ensuremath{I}}
\newcommand{\optset}{\ensuremath{O}}
\newcommand{\nodeset}{\ensuremath{N}}
\newcommand{\tailcost}{\ensuremath{c_T}}
\newcommand{\regencost}{\ensuremath{c_R}}
\newcommand{\portcost}{\ensuremath{c_{P}}}
\newcommand{\regendist}{\t{\sc regen\_dist}}

\newcommand{\srcip}{\ensuremath{\alpha}}
\newcommand{\dstip}{\ensuremath{\beta}}
\newcommand{\srcipnode}{\ensuremath{a}}
\newcommand{\dstipnode}{\ensuremath{b}}
\newcommand{\node}{\ensuremath{u}}
\newcommand{\router}{\node}
\newcommand{\srcopt}{\node}
\newcommand{\dstopt}{\ensuremath{v}}
\newcommand{\tap}{\ensuremath{T'}}
\newcommand{\rap}{\ensuremath{R'}}

\newcommand*{\thead}[1]{\multicolumn{1}{c}{\bfseries #1}}

\renewcommand{\checkmark}{{\color{OliveGreen}\ding{52}}}
\newcommand{\xmark}{\text{\color{red}\ding{55}}}

\begin{document}
\title{Robust Network Design for Software-Defined IP/Optical Backbones}

\author{Jennifer~Gossels,
        Gagan~Choudhury,
        and~Jennifer~Rexford\thanks{J. Gossels and J.Rexford are with the Computer Science Department of
Princeton University, Princeton, NJ.}\thanks{G. Choudhury is with AT\&T Labs Research, Middletown, NJ.}}

\maketitle

\begin{abstract}
Recently, Internet service providers (ISPs) have gained increased flexibility in how
they configure their in-ground optical fiber into an IP network.  This greater
control has been made possible by
\begin{inparaenum}[(i)]
	\item the maturation of software defined networking (SDN), and
	\item improvements in optical switching technology.
\end{inparaenum}
Whereas traditionally, at network design time,
each IP link was assigned a fixed optical path and bandwidth, modern colorless
and directionless Reconfigurable Optical Add/Drop Multiplexers (CD ROADMs)
allow a remote SDN controller to remap the IP topology to the optical underlay
on the fly.
Consequently, ISPs face new opportunities and challenges in
the \emph{design} and \emph{operation} of their backbone networks
\cite{Birk2016, Choudhury2017, Choudhury2018, Tse2018}.

Specifically, ISPs must determine how best
to design their networks to take advantage of the new capabilities;
they need an automated way to generate the least expensive network design that
still delivers all offered traffic, even in the presence of equipment failures.  This
problem is difficult because of the physical constraints governing the placement of
optical regenerators, a piece of optical equipment necessary for maintaining an
optical signal over long stretches of fiber.
As a solution, we present an integer linear program (ILP) which
\begin{inparaenum}[(1)]
	\item solves the equipment-placement network design problem;
	\item determines the optimal mapping of IP links to the optical infrastructure
	for any given failure scenario; and
	\item determines how best to route the offered traffic over the IP topology.
\end{inparaenum}
To scale to larger networks,
we also describe an efficient heuristic
that finds nearly optimal network designs in a fraction of the time.
Further, in our experiments 
our ILP offers cost savings of up to 29\% compared to traditional
network design techniques. \end{abstract}

\section{Introduction}
\label{sec:intro}

\begin{figure}
\includegraphics[width=\columnwidth]{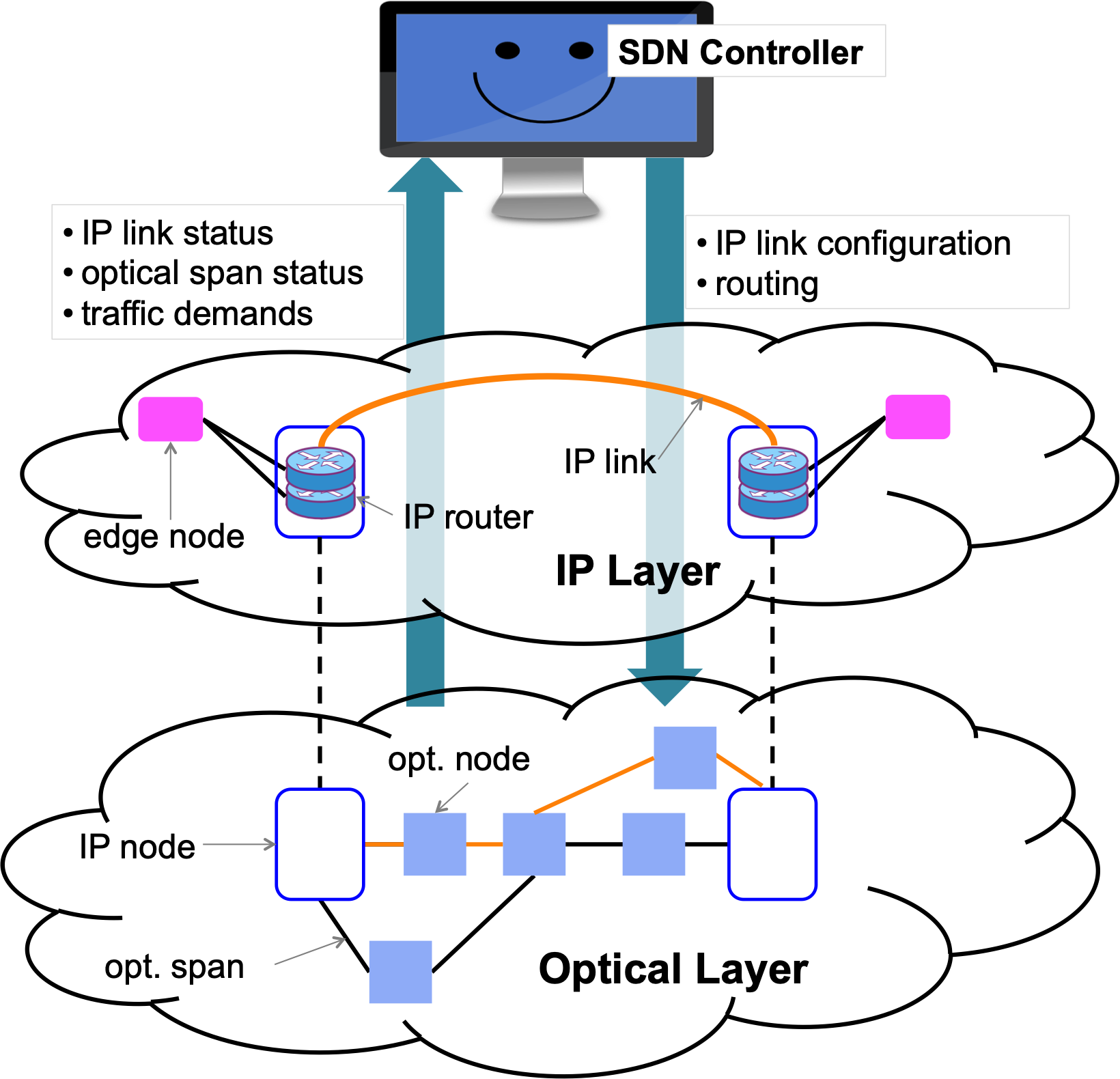}
\caption{Layered IP/optical architecture.
The highlighted orange optical spans comprise one possible mapping of the
orange IP
link to the optical layer.  Alternatively, the SDN controller could remap
the same orange IP link to follow the black optical path.}
\label{fig:layered-architecture}
\end{figure}

Over the past several years, the advent of software defined networking (SDN),
along with improvements in optical switching technology,
has given network operators more flexibility in configuring their in-ground
optical fiber into an IP network.  Whereas traditionally, at network design time,
each IP link was assigned a fixed optical path and bandwidth, modern SDN
controllers can program colorless
and directionless Reconfigurable Optical Add/Drop Multiplexers (CD ROADMs)
to remap the IP topology to the optical underlay on the fly, while the
network continues carrying traffic and without deploying technicians to
remote sites (Figure \ref{fig:layered-architecture}) \cite{Birk2016, Choudhury2017, Choudhury2018, Tse2018}.

In the traditional setting, if a router failure or fiber cut causes an IP
link to go down, all resources that
were being used for said IP link are rendered useless.  There are two
viable strategies to recover from any single optical span or IP router
failure.  First, we could independently restore the optical and
IP layers, depending on the specific failure; we could perform pure optical recovery
in the case
of an optical span failure or pure IP recovery in the case of an
IP router failure.  Note that the strategy we refer to as ``pure optical recovery''
of course involves reestablishing the IP link over the new optical path.  We call it
``pure optical recovery'' because once the link has been recreated over the new optical
path, the change is transparent to the IP layer.
Second, we could design the network with sufficient capacity and path diversity
that we can at runtime perform pure IP restoration.  In practice, ISPs have used
the latter strategy, as it is generally more resource efficient \cite{Chiu2001}.

Now, the optical
and electrical equipment can be repurposed for setting up
the same IP link along a different path, or even for setting up a
different IP link.  In the context of failure recovery, the important upshot
is that joint multilayer (IP and optical) failure recovery is now possible at runtime.
The SDN controller is responsible for performing this remote
reprogramming of both CD ROADMs and routers; while we generally think of SDN
as operating at the network layer and above, it is now extending into the physical layer.

\begin{figure}
\includegraphics[width=\columnwidth]{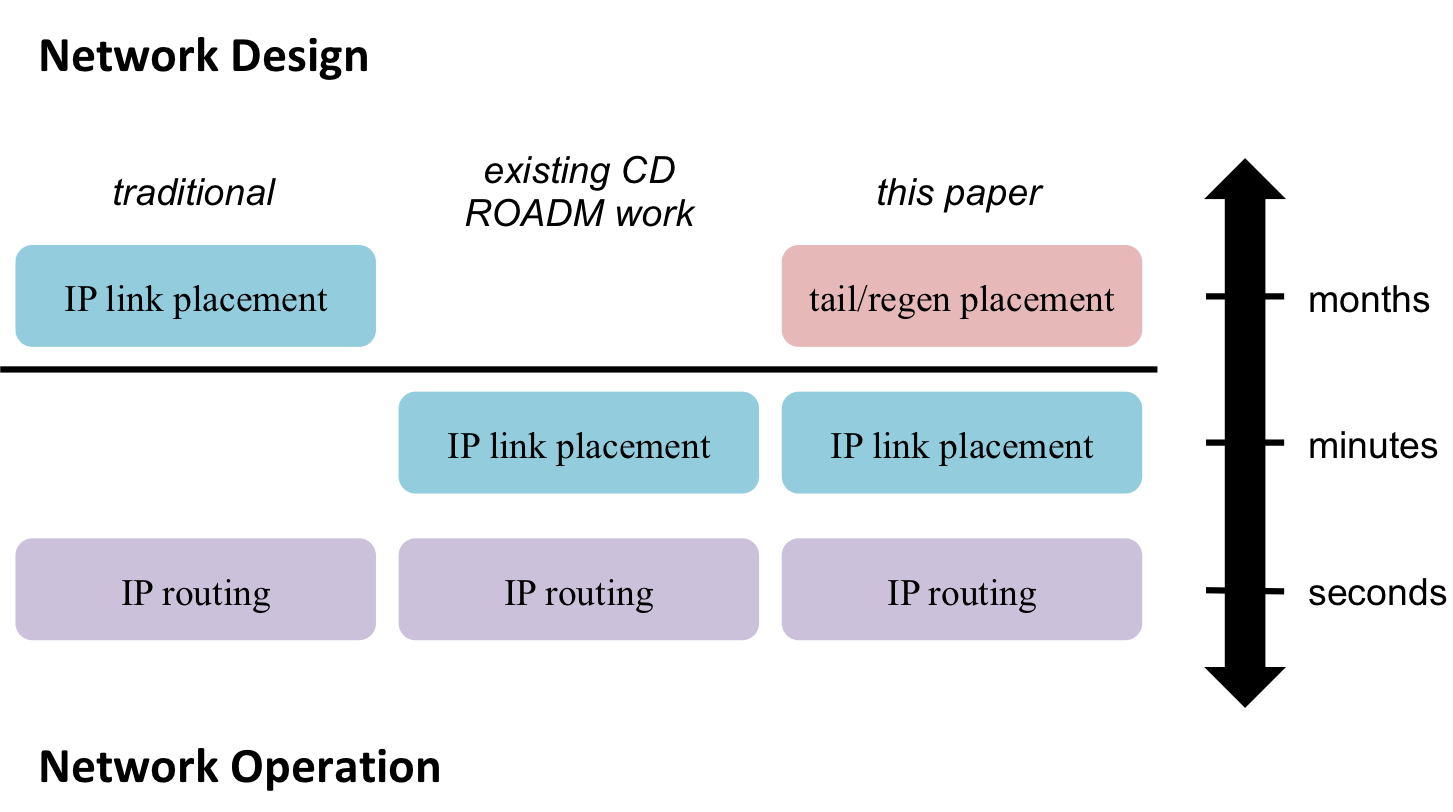}
\caption{Components of network design vs. network operation in, from left
to right: traditional networks, existing studies on how best to take advantage
of CD ROADMs, and this paper.  The vertical dimension is timescale.}
\label{fig:design-vs-operation}
\end{figure}

Thus, SDN-enabled
CD ROADMs
shift the boundary between network design and network operation
(Figure \ref{fig:design-vs-operation}).
We use the term network \emph{design} to refer to any changes that happen on a
human timescale, e.g., installing new routers or dispatching a crew to fix a failed link.
We use network \emph{operation} to refer to changes that can happen on a smaller
timescale, e.g., adjusting routing in response to switch or link failures or changing
demands.

\begin{figure*}
\centering
\includegraphics[width=0.8\textwidth]{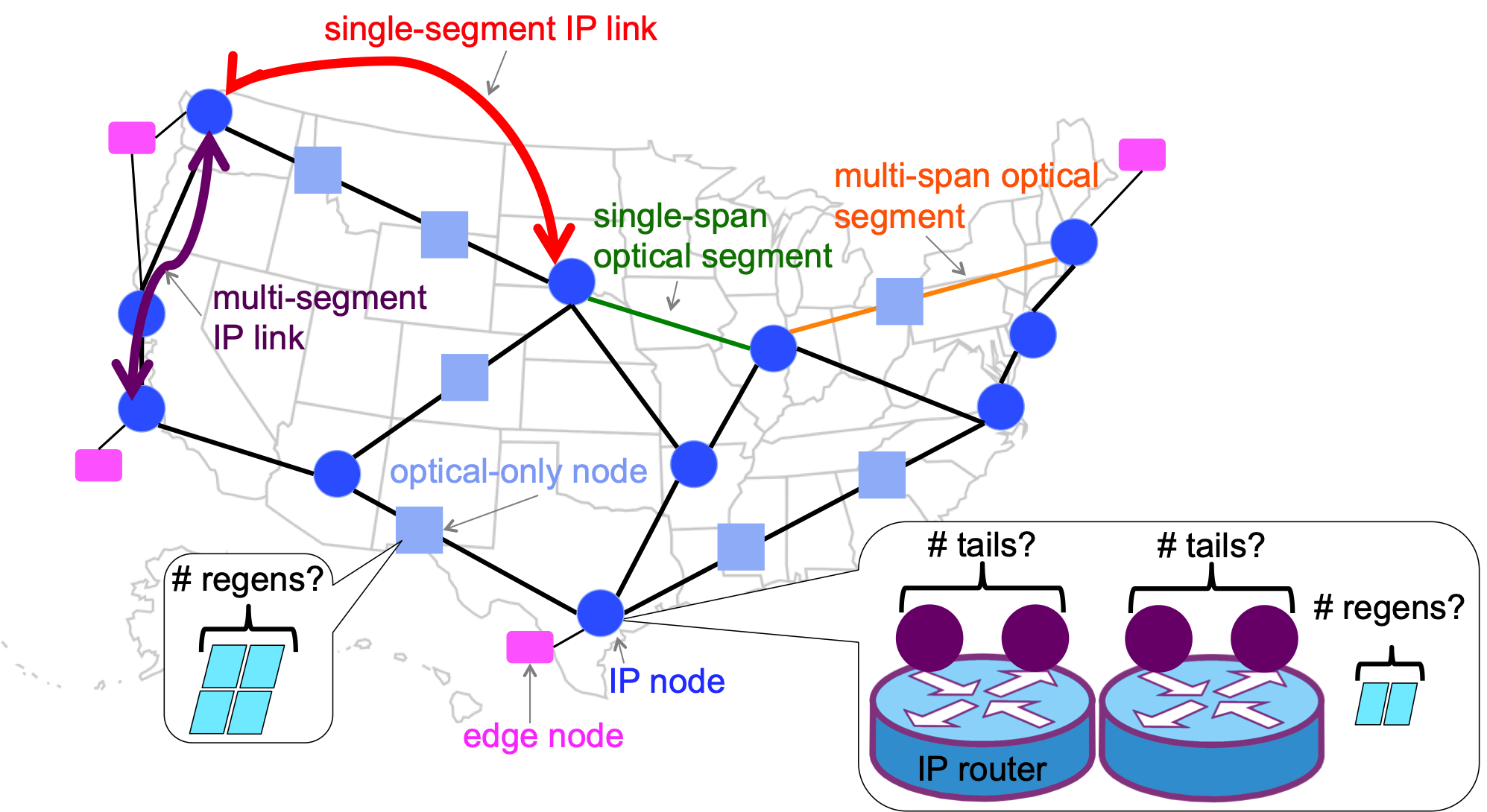}
\caption{IP/optical network terminology.}
\label{fig:backbone}
\end{figure*}

As Figure \ref{fig:design-vs-operation} shows, network design \emph{used
to} comprise IP link placement.  To describe what it now entails,
we must provide background on
IP/optical backbone architecture (Figure \ref{fig:backbone}).  The limiting
resources in the design of an IP backbone are the equipment
housed at each IP and optical-only node.
Specifically, an IP node's responsibility is to terminate optical links and
convert the optical signal to an electrical signal, and to do so it needs
enough \emph{tails} (\emph{tail} is shorthand for the combination of an
optical transponder and a router port).  An optical node must
maintain the optical signal over long distances, and to do so it needs
enough \emph{regenerators} or \emph{regens} for the IP links passing through it.
Therefore, we precisely state the new network design problem as follows:
\emph{Place tails
and regens in a manner that minimizes cost while allowing the network to
carry all expected traffic, even in the presence of equipment failures}.

This new paradigm creates both new opportunities and challenges in
the design and operation of backbone networks \cite{Chiu2007}.
Previous work has explored the advantages of joint multilayer optimization
over traditional IP-only optimization
\cite{Birk2016, Choudhury2017, Choudhury2018, Tse2018}
(e.g., see Table 1 of \cite{Choudhury2018}).  However, these authors
primarily resorted to heuristic optimization and
restoration algorithms, due to the restrictions of routing (avoiding splitting
flows into arbitrary proportions), the need for different restoration and latency
guarantees for different quality-of-service classes, and the desirability of fast
run times.

Further complicating matters is that network
components fail, and when they do a production backbone must reestablish
connectivity within seconds.  Tails and regens
cannot be purchased or relocated at this timescale, and therefore our network
design must be \emph{robust} to a set of possible failure scenarios.  Importantly,
we consider as \emph{failure scenarios} any single optical fiber cut or IP router failure.
There are other possible causes of failure (e.g., single IP router port, ROADM,
transponder, power failure), which allow for various alternative recovery
techniques, but we focus on these two.
Existing techniques respond efficiently to IP layer failures \cite{Chiu2007} \emph{or}
optical layer failures, but ours is the first to jointly optimize over the two.

Thus, we overcome three main challenges to
present an exact formulation and solution to the network design problem.
\begin{enumerate}
	\item The solution must be a single tail and regen configuration
	that works for all single IP router and optical fiber failures. This configuration
	should
	minimize cost
	under the assumption that the IP link topology will be reconfigured in
	response to each failure. \label{challenge1}
	\item The positions of regens relative to each other along the optical path
	determine
	which IP links are possible. \label{challenge2}
	\item The problem is computationally complex because it
	requires integer variables and constraints.  Each
	tail and each regen supports a 100 Gbps IP link.  Multiple
	tails or multiple regens can be combined at a single location to
	build a faster link, but they can't be split into e.g., 25 Gbps units
	that cost 25\% of a full element.
\end{enumerate}

These challenges arise because the recent shift in the boundary
between network design and operation
fundamentally changes the design problem; simply including
link placement in network operation optimizations does not suffice to
fully take advantage of CD ROADMs. A network design is
optimal relative to a certain set of assumptions about what can be
reconfigured at runtime. Hence, traditional network designs are
only optimal \emph{under the assumption that tails and regens
are fixed to their assigned IP links}.
With CD ROADMs, the optimal network design must be computed
\emph{under the assumption that IP links will be adjusted} in response
to failures or changing traffic demands.

To this end, we make three main contributions.
\begin{enumerate}
	\item After describing the importance of jointly optimizing over
	the IP and optical layers in Section \ref{sec:background}, we formulate
	the optimal network design algorithm
	(Section \ref{sec:problem}).  In this way we address
	challenges \eqref{challenge1} and \eqref{challenge2} from above.
	\item We present two scalable, time-efficient approximation algorithms
	for the network design problem, addressing the computational
	complexity introduced by the integer constraints (Section
	\ref{sec:scalable approximations}), and we explain which use cases
	are best suited to each of our algorithms (Section \ref{subsec:roles}).
	\item We evaluate our three algorithms in relation to each other and to
	legacy networks (Section \ref{sec:eval}).
\end{enumerate}
We discuss related work in Section \ref{sec:related} and conclude in Section
\ref{sec:conclusion}.
 \section{IP/Optical Failure Recovery}
\label{sec:background}
In this section we provide more background IP/optical networks.
We begin by defining key terms
and introducing a running example (Section
\ref{subsec:background terminology example}).  We then
use this example to discuss various failure recovery options in
both traditional (Section \ref{subsec:traditional}) and CD ROADM
(Section \ref{subsec:background CD ROADM})
IP/optical networks.

\subsection{IP/Optical Network Architecture}
\label{subsec:background terminology example}
As shown in Figure \ref{fig:backbone}, an IP/optical network consists
of optical fiber,
the IP nodes where fibers meet, the optical nodes stationed intermittently
along fiber segments, and the edge nodes that serve as the sources and
destinations of traffic.  We do not consider the links
connecting an edge router to a core IP router as part of our design problem; we
assume these are already placed and fault tolerant.

Each IP node houses one
or more IP \emph{routers}, each with zero or more tails, and zero or more
optical regens.  Each optical-only node houses zero or more optical
regens but cannot contain any routers (Figure
\ref{fig:backbone}).
While IP and optical nodes serve
as the endpoints of optical spans and segments,
specific IP routers serve as the
endpoints of IP links.

For our purposes, an \emph{optical span} is the smallest unit describing a stretch
of optical fiber; an optical span is the section of fiber between
any two nodes, be they IP or optical-only.  Optical-only nodes can join multiple
optical spans into a single \emph{optical segment}, which
is a stretch of fiber terminated at both ends by IP nodes.  The path of a single
optical segment may contain one or more optical-only nodes.  The physical
layer underlying each \emph{IP link} comprises one or more
optical segments.
An IP link is terminated at each end by a specific IP router and can travel
over multiple optical segments if its path traverses an intermediate IP node
without terminating at one of that node's routers.  Figure
\ref{fig:backbone}
illustrates the roles of optical spans and segments and IP links.
The locations of all nodes and optical spans
are fixed and cannot be changed, either at design time or during network
operation.  

An optical signal can travel only a finite distance
along the fiber before it must be regenerated; every \regendist\ miles
the optical signal must
pass through a regen, where it is converted from an optical signal to an electrical
signal and then back to optical before being sent out the other end.  The exact
value of \regendist\ varies depending on the specific optical components, but
it is roughly 1000 miles for our setting of a long-distance ISP backbone with
100 Gbps technology.  We use the value of $\regendist = 1000$ miles throughout
this paper.

\begin{figure}
\includegraphics[width=\columnwidth]{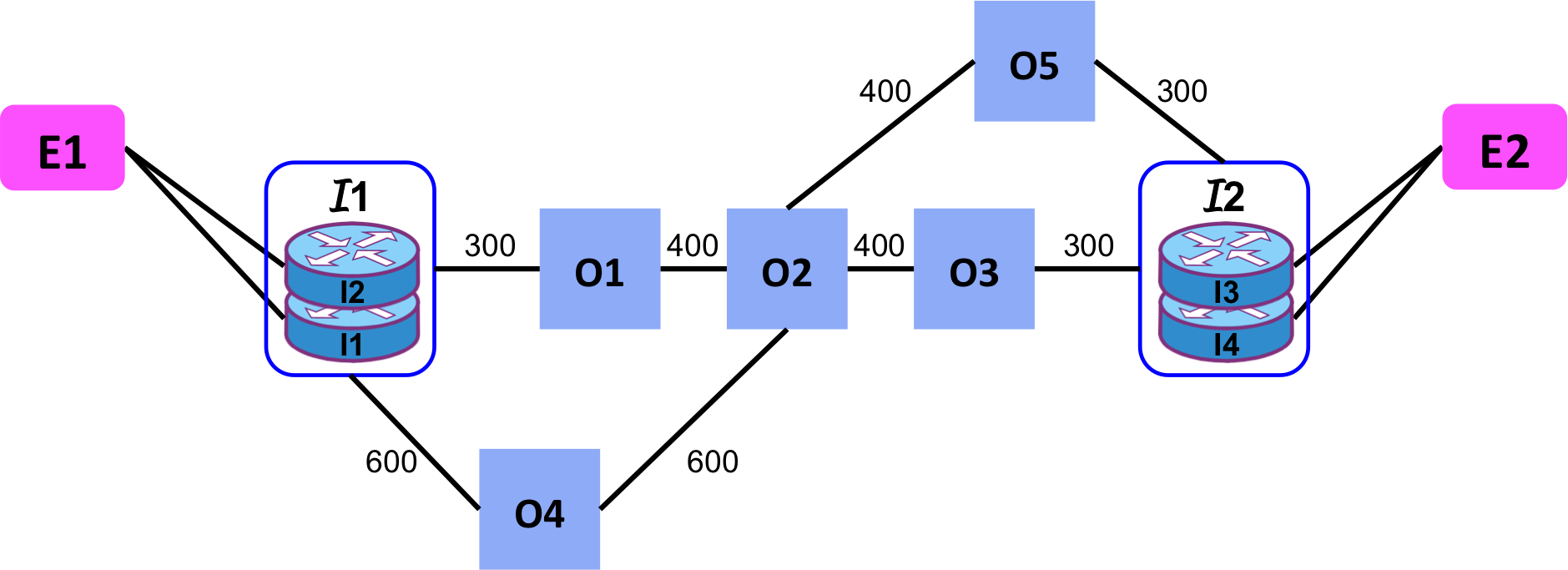}
\caption{Example optical network illustrating the different options for
failure restoration.  The
number near each edge is the edge's length in miles.}
\label{fig:background-example}
\end{figure}

\para{Example network design problem.}
The network in Figure \ref{fig:background-example}
has two IP nodes, \ipnodename{1} and \ipnodename{2}, and five optical-only nodes,
\optname{1}-\optname{5}.  \ipnodename{1} and \ipnodename{2}
each have two IP routers (\routername{1}, \routername{2} and \routername{3},
\routername{4}, respectively).  Edge routers \edgename{1} and \edgename{2}
are the sources and destinations of all traffic.  The problem is to design the optimal IP
network, requiring the
fewest tails and regens, to carry 80 Gbps from \edgename{1} to \edgename{2} while
surviving any single optical
span or IP router failure.  We do not consider failures of \edgename{1} or \edgename{2},
because
failing the source
or destination would render the problem trivial or impossible, respectively.

If we don't need to be robust to any failures, the optimal solution is to add one 100 Gbps
IP link from \routername{1} to \routername{3} over the nodes \ipnodename{1},
\optname{1}, \optname{2}, \optname{3}, and \ipnodename{2}.  This solution requires
one tail each at \routername{1} and \routername{3} and one regen at \optname{2},
for a total of two tails and one regen.

\subsection{Failure Recovery in Traditional Networks}
\label{subsec:traditional}
In the traditional setting, the design problem is to place IP links;
in this setting, once an IP link is placed at
design time, its tails and regens are permanently committed to it.
If one optical span or router
fails, the entire IP link fails and the rest of its resources lie idle.
During network operation, we may only adjust routing over the established
IP links.

In general, this setup allows for four possible types of failure restoration.
Two of these techniques are inadequate because they cannot recover from
all relevant failure scenarios (first two rows of Table \ref{tab:failure restoration}).
The other two are effective but suboptimal in their resource requirements (second
two rows of Table \ref{tab:failure restoration}).  We describe these four approaches
below, guided by the running example shown in Figure \ref{fig:background-example}.
In Section \ref{subsec:background CD ROADM} we show that CD ROADMs
allow for a network design
that meets our problem's requirements in a more cost-effective way.

\begin{table}
\caption{Properties of various failure recovery approaches.  The first four techniques
are possible in legacy and CD ROADM networks, while the fifth requires CD ROADMs.}
\label{tab:failure restoration}
\begin{tabular}{l|cccc}
\multicolumn{1}{c|}{\textbf{Recovery Technique}} &
\thead{\# Tails} & \thead{\# Regens} & \thead{IP?} & \thead{Optical?} \\
\hline
pure optical & 2 & 2 & \xmark & \checkmark \\
pure IP, shortest path & 4 & 4 & \checkmark & \xmark \\
pure IP, any path & 4 & 3 & \checkmark & \checkmark \\
separate IP and optical & 4 & 4 & \checkmark & \checkmark \\[0.3em]
\textbf{joint IP/optical} & \textbf{4} & \textbf{2} & \checkmark & \checkmark
\end{tabular}
\end{table}

\para{Inadequate recovery techniques.}
In \emph{pure optical layer} restoration, if an optical span fails, we reroute each
affected IP link over the optical network
by avoiding the failed span.  The rerouted path may require additional regens.
In the example shown in Figure \ref{fig:background-example}, this amounts to
rerouting the IP link along the alternate path
\ipnodename{1}-\optname{4}-\optname{2}-\optname{5}-\ipnodename{2}
whenever any optical span fails.
This path requires one regen each at \optname{4} and \optname{2}. However,
because the (\ipnodename{1}, \ipnodename{2})
link will never be instantiated over both paths simultaneously, the second path
can reuse the original regen \optname{2}.  Hence, we need only buy one extra regen at
\optname{4}, for a total of two tails (at \ipnodename{1} and \ipnodename{2}) and
two regens (at \optname{2} and \optname{4}).  The problem
with this pure optical restoration strategy is that it cannot protect against IP router
failures.

In \emph{pure IP layer restoration with each IP link routed along its shortest
optical path}, we maintain enough fixed IP links such that during any failure condition,
the surviving IP links can carry the required traffic.  If any component
of an IP link fails, then the entire IP link fails and even the intact components cannot be
used.  In large networks, this policy usually finds a feasible solution to protect against
any single router or optical span failure.  However, it may not be optimally cost-effective
due to the restriction that
IP links follow the shortest optical paths.  Furthermore, in small networks it
may not provide a solution that is robust to all optical span failures.

If we only care about IP layer failures, the optimal strategy for our running example
is to place two 100 Gbps
links, one from \routername{1} to \routername{3} and a second from \routername{2}
to \routername{4} and both following the optical path
\ipnodename{1}-\optname{1}-\optname{2}-\optname{3}-\ipnodename{2}.
Though this design is robust to the
failure of any one of \routername{1}, \routername{2}, \routername{3}, and \routername{4},
it cannot protect against optical span failures.

\para{Correct but suboptimal recovery techniques.}
In contrast to the two failure recovery mechanisms described above, the following
two techniques can correctly recover from any single IP router or optical span failure.
However, neither reliably produces the least expensive network design.

\emph{Pure IP layer restoration with no restriction on how IP links are routed
over the optical network} is the same as IP restoration over shortest paths except
IP links can be routed over any optical path.  With this policy, we always
find a
feasible solution for all failure conditions, and it finds the most cost-effective
among the possible pure-IP solutions.
However, its solutions still require more tails or regens than those produced by our 
ILP, and solving for this
case is computationally complex.
In terms of Figure \ref{fig:background-example}, pure IP restoration with no restriction
on IP links' optical paths entails routing the (\routername{1}, \routername{3}) IP link along the
\ipnodename{1}-\optname{1}-\optname{2}-\optname{3}-\ipnodename{2} path
and the (\routername{2}, \routername{4}) IP link along the
\ipnodename{1}-\optname{4}-\optname{2}-\optname{5}-\ipnodename{2} path.
This requires two tails
plus one regen (at \optname{2}) for the first IP link and two tails plus two regens (at
\optname{4} and \optname{2})
for the second IP link, for a total of four tails and three regens.

The final failure recovery technique possible in legacy networks, without CD ROADMs,
is \emph{pure IP layer restoration for router failures and pure optical layer restoration
for optical failures.}
This policy works in all cases but is usually more expensive
than the two pure IP layer restorations mentioned above.  In terms of our running
example, we need two tails and two regens for each of two IP links, as we showed
in our discussion of pure IP recovery along shortest paths. Hence,
this strategy requires a total of four tails and four regens.

In summary, the optimal network design with legacy technology that is robust to optical
and IP failures requires four tails and three regens.

\subsection{Failure Recovery in CD ROADM Networks}
\label{subsec:background CD ROADM}
A modern IP/optical network architecture is identical to that described in Section
\ref{subsec:background terminology example} aside from the presence of an SDN
controller.  This single logical controller
receives notifications of the changing status of any IP or optical
component and also any changes in traffic demands between any pair of edge routers
and uses this information compute the optimal IP link configuration and the optimal
routing of traffic over these links.  It then communicates the relevant link configuration
instructions to the CD ROADMs and the relevant forwarding table changes to the
IP routers.

As in the traditional setting, we cannot add or remove edge nodes, IP nodes, optical-only
nodes, or optical fiber.  But, now the design problem is to decide how many tails to place
on each router and how many regens to place at each IP and optical node;
no longer must we commit to fixed IP links at design time.
Routing remains a key component of the network design problem, though it is now
joined by IP link placement.

Any of the four existing failure recovery techniques is possible in a modern network.
In addition, the presence of SDN-controlled CD ROADMs
allows for a fifth option, joint IP/optical recovery.  In contrast to the traditional setting,
IP links can now be reconfigured at runtime.  As above, suppose
the design calls for an IP link between routers \ipnodename{1} and \ipnodename{2}
over the optical
path \routername{1}-\routername{2}-\routername{3}-\routername{4}.
Now, these resources are \emph{not} permanently committed
this IP link.  If one component
fails, the remaining tails and regens can be repurposed either to reroute the
(\ipnodename{1}, \ipnodename{2})
link over a different optical path or to (help) establish an entirely new IP link.

Returning to our running example, with joint IP/optical restoration, we can recover from
any single IP or optical failure with just one IP link from \routername{1} to \routername{3}.
If there is any optical link failure
then this link shifts from its original shortest path, which needs a regen at O2, to the path
\ipnodename{1}-\optname{4}-\optname{2}-\optname{5}-\ipnodename{2}, which needs regens
at \optname{2} and \optname{4}.  Importantly, the regen at \optname{2} can be reused.
Hence, thus far we need two tails and two regens.  To account for the possibility of \routername{1}
failing, we add an extra tail at \routername{2}; if \routername{1} fails then at runtime we create an IP
link from \routername{2} to \routername{3} over the path \ipnodename{1}-\optname{1}-\optname{2}-\optname{3}-\ipnodename{2}.
Since this link is only active in the case
that \routername{1} has failed, it will never be instantiated at the same time as the
(\routername{1}, \routername{3}) link
and can therefore
reuse the regen we already placed at \optname{2}.  Finally, to account for the possibility
of \routername{3} failing,
we add an extra tail at \routername{4}.  This way, at runtime we can create the IP link
(\routername{1}, \routername{4}) along the path
\ipnodename{1}-\optname{1}-\optname{2}-\optname{3}-\ipnodename{2}.
Again, only one of these IP links will ever be active at one time,
so we can reuse the regen at \optname{2}.  Therefore, our final joint optimization design requires
four tails and two regens.  Hence, even in this simple topology, compared to the most
cost efficient traditional strategy, joint IP/optical optimization and failure recovery
saves the cost of one regen.

\subsubsection{A note on transient disruptions}
As shown in Figure \ref{fig:design-vs-operation}, IP link configuration operates on the order of minutes, while routing operates on sub-second timescales.  IP link configuration takes several minutes because the process entails the following three steps:
\begin{enumerate}
	\item \label{reconfiguration1}
	Adding or dropping certain wavelengths at certain ROADMs;
	\item \label{reconfiguration2}
	Waiting for the network to return to a stable state; and
	\item \label{reconfiguration3}
	Ensuring that the network is indeed stable.
\end{enumerate}
A ``stable state'' is one in which the optical signal reaches tails at IP link endpoints with
sufficient optical power to be correctly converted back into an electrical signal.
Adding or dropping wavelengths at ROADMs temporarily reduces the signal's power
enough to interfere with this optical-electrical conversion, thereby rendering the
network temporarily unstable.  Usually, the network correctly returns to a stable state
within seconds of reprogramming the wavelengths (i.e., steps \eqref{reconfiguration1} and
\eqref{reconfiguration2} finish within seconds).  However, to ensure that the network is
always operating with a stable physical layer (step \eqref{reconfiguration3}), manufacturers
add a series of tests and adjustments to the reconfiguration procedure.  These tests take
several minutes, and therefore step \eqref{reconfiguration3} delays completion of the
entire process. Researchers are currently working to bring reconfiguration
latency down to the order of milliseconds \cite{Chiu2012}, similar to the timescale at which
routing currently operates.  However, for now we must account for a transition period of
approximately two minutes
when the link
configuration has not yet been updated and is therefore not optimal for the new
failure scenario.

During this transient period, the network
may not be able to deliver all the offered traffic.  We mitigate this harmful traffic loss
by immediately reoptimizing routing over the existing topology while the network is
transitioning to its new configuration.  As we show in Section \ref{subsec:transient},
by doing so we
successfully deliver the vast majority of offered traffic under almost
all failure scenarios.  Many operational ISPs carry multiple classes of traffic,
and their service level agreements (SLAs) allow them to drop some low priority
traffic under failure or extreme congestion.  At one large
ISP, approximately 40-60\% of traffic is low priority.  We always deliver
at least 50\% of traffic just by rerouting. \section{Network Design Problem}
\label{sec:problem}
We now describe the variables and constraints of our integer linear
program (ILP) for solving the network design problem.
After formally stating the objective function in Section
\ref{subsec:objective} we introduce the problem's constraints
in \ref{subsec:tails regens} and \ref{subsec:IP links}.
To avoid cluttering our presentation of the main ideas of the
model, throughout \ref{subsec:objective} -
\ref{subsec:IP links} we assume exactly one router per IP
node.  In \ref{subsec:extensions} we relax this assumption,
and we also explain how to extend the model to changing
traffic demands.

For ease of explanation, we elide the distinction between edge
nodes and IP nodes; we treat IP nodes as the ultimate sources
and destinations of traffic.

\subsection{Minimizing Network Cost}
\label{subsec:objective}
Our inputs are
\begin{inparaenum}[(i)]
	\item the optical topology, consisting of the set \ipnodeset\ of
	IP nodes, the set \optset\ of optical nodes, and the fiber
	links (annotated with distances) between them; and
	\item the demand matrix $D$.
\end{inparaenum}

We use the variable $T_{\router}$ to represent the number of tails
that should be placed at router $\router$, and $R_{\node}$ represents the number of
regens at node \node.  An optical-only node can't have any tails.

The capacity of an IP link $\ell = (\srcip, \dstip)$ is limited by the number
of tails dedicated to $\ell$ at \srcip\ and \dstip\ and the number of
regens dedicated to $\ell$.  Technically,
the original signal emitted by \srcip\ is strong enough to travel \regendist, and
$\ell$ doesn't need regens there.  However, for ease of explanation, we assume that
$\ell$ does need regens at \srcip, regardless of its length.
This requirement of regens at the beginning of each IP link is necessary only for the
mathematical model and not in the actual network.  We add a
trivial postprocessing step to remove these regens from the final
count before reporting our results.
Table \ref{tab:notation}
summarizes our notation.

\begin{table*}
\centering
\caption{Notation.}
\label{tab:notation}
\begin{tabular}{|c|l|l|}
\hline
& & \multicolumn{1}{c|}{\textbf{Definition}} \\
\hline
\multirow{5}{*}{\textbf{Inputs}} &
\ipnodeset & set of IP nodes \\
& \routerset & set of IP routers \\
& \optset & set of optical-only nodes \\
& \nodeset & set of all nodes ($\nodeset = \ipnodeset \cup \optset$) \\
& $D$ & demand matrix, where $D_{st} \in D$ gives the demand from
IP node $s$ to IP node $t$ \\
& $F$ & set of all possible failure scenarios $F = \{f_1, f_2, \dots, f_n\}$ \\
& $dist_{\srcopt\dstopt f}$ & shortest distance from optical node \srcopt\
to optical node \dstopt\ in failure scenario $f$ \\
& $\optset_{\node f}$ & set of all next-hops \dstopt\ with
$dist_{\node\dstopt f} < \regendist$ \\

\hline

\textbf{Outputs} &
$T_\router$ & number of tails placed at IP router \router \\
{\small(Network Design)} & $R_\node$ & total regens placed at optical node \node\\[0.75em]
\textbf{Outputs} &
$X_{\srcip\dstip f}$ & capacity of IP link $(\srcip, \dstip)$ in failure scenario $f$  \\
{\small(Network Operation)} & $Y_{st\srcip\dstip f}$ & amount of $(s, t)$ traffic routed on IP link $(\srcip, \dstip)$
in failure scenario $f$ \\

\hline

\textbf{Intermediate} & $R_{\srcip\dstip\srcopt\dstopt f}$
& number of regens at \srcopt\ for
optical segment $(\srcopt, \dstopt)$ of IP link $(\srcip, \dstip)$ in failure $f$ \\
\textbf{Values} & $R_{\node f}$
& number of regens needed at optical node \node\ in failure scenario $f$ \\
\hline
\end{tabular}
\end{table*}

Our objective is to place tails and regens to minimize the ISP's
equipment costs while ensuring that the network can carry all necessary
traffic under all failure scenarios.  Let \tailcost\ and \regencost\ be the cost
of one tail and one regen,
respectively.  Then the total cost of all tails is
$\tailcost \sum_{\router \in \routerset} T_{\router}$,
the total cost of all regens is $\regencost \sum_{\node \in \optset} R_\node$,
and our objective is
\begin{equation*}
\min~~ \tailcost \sum_{\router \in \routerset} T_{\router} +
\regencost \sum_{\node \in \optset} R_\node.
\end{equation*}

The stipulation that the output tail and regen placement work for all
failure scenarios is crucial.  Without some dynamism in the inputs, be it
from a changing topology across failure scenarios or from a changing
demand
matrix, CD ROADMs' flexible reconfigurability would be useless.  We
focus on robustness to IP router and optical span failures because conversations with
one large ISP indicate that failures affect network conditions
more than routine demand fluctuations.  Extending our model to
find a placement robust to both equipment failures and changing demands
should be straightforward.

\subsection{Robust Placement of Tails and Regens}
\label{subsec:tails regens}
In traditional networks, robust design requires choosing a single IP link
configuration that is optimal for all failure scenarios under the assumption
that routing will depend on the specific failure state \cite{Chiu2007}.
With CD ROADMs,
robust network design requires choosing a single tail/regen placement
that is optimal for all failure scenarios under the assumption that
both routing and the IP topology will depend on the specific failure state.
In either case, solving the network design problem requires solving
the network operation problem as an ``inner loop''; to determine the optimal
network design we need to simulate how a candidate network would
operate, in terms of IP link placement and routing, in each failure scenario.

At the mathematical level, CD ROADMs introduce two additional sets of
decision variables to the traditional network design
optimization.  With the old technology, the problem is to
optimize over two sets of decision variables:  one set
for where to place IP links and
what the capacities of those links should be, and a second set for which
links different volumes of traffic should traverse.
In traditional network design, there is no need to explicitly model
tails and regens separate from link placement, because each tail or
regen is associated with
exactly one IP link.
Now, any given tail or regen is not associated with exactly one IP link.
Thus, we must decide not only link placement and routing but also
the number of tails to place at each IP node and the number of regens
to place at each site.  We describe these two novel aspects of our
formulation in turn.

\para{Constraints governing tail placement.}
Our first
constraint requires that the number of tails placed at any
router \router\ is enough to accommodate all the IP links \router\
terminates:
\begin{eqnarray}
\sum_{\srcip \in \routerset} X_{\srcip \router f} & \leq & T_{\router} 
\label{eq:t>=incoming} \\
\sum_{\dstip \in \routerset} X_{\router\dstip f} & \leq & T_{\router}
\label{eq:t>=outgoing} \\[-0.5em]
& & \forall \router \in \routerset, \forall f \in F \nonumber
\end{eqnarray}
As shown in Table \ref{tab:notation}, $X_{\srcip\router f}$ is the capacity
of IP link $(\srcip, \router)$ in failure scenario $f$.  Hence,
$\sum_{\srcip \in \routerset} X_{\srcip \router f}$ is the total incoming
bandwidth terminating at router \router, and Constraint \eqref{eq:t>=incoming}
says that \router\ needs at least this number of tails.  Analogously,
$\sum_{\dstip \in \routerset} X_{\router\dstip f}$ is the total
outgoing bandwidth from \router, and Constraint \eqref{eq:t>=outgoing}
ensures that \router\ has enough tails for these links, too.  We don't need
$T_{\router}$ greater than the sum of these quantities because each
tail supports a bidirectional link.

\para{Constraints governing regen placement.}
The second fundamental difference between our model and existing work is that
we must account for relative positioning of regens both within and across
failure scenarios.  Because of physical limitations in the distance an optical
signal can travel,
no IP link can include a span longer than \regendist\
without passing through a regenerator.  As a result, the decision to place
a regen at one optical location depends on the decisions we make about
other locations, both within a single failure scenario and across changing
network conditions.  Therefore, we introduce auxiliary variables
$R_{\srcip\dstip\node\dstopt f}$ to represent the number of regens
to place at node \node\ for the link between IP routers $(\srcip, \dstip)$ in failure
scenario $f$ \emph{such that the next regen traversed will be at node \dstopt}.

Ultimately, we want to solve for $R_\node$, the
number of regens to place at \node, which doesn't depend on the IP link,
next-hop regen, or failure scenario.  But, we need the
$R_{\srcip\dstip\node\dstopt f}$ variables to encode these dependencies
in our constraints.  We connect $R_\node$ to $R_{\srcip\dstip\node\dstopt f}$
with the constraint
\begin{equation}
\label{eq:r>=outgoing}
R_{\node} \geq \sum_{\substack{\srcip, \dstip \in \routerset \\ \dstopt \in \optset}}
R_{\srcip\dstip\node\dstopt f} ~~ \forall \node \in \optset, \forall f \in F.
\end{equation}

We use four additional constraints for the $R_{\srcip\dstip\node\dstopt f}$
variables.  First, we prevent some node \dstopt\ from being
the next-hop regen for some node \srcopt\ if the shortest path
between \srcopt\ and \dstopt\ exceeds \regendist:
\begin{eqnarray*}
\label{eq:regens constr last}
R_{\srcip\dstip\srcopt\dstopt f} & =  & 0 \\
& & \forall \srcip, \dstip \in \routerset, \nonumber \\[-0.5em]
& & \forall \srcopt, \dstopt~\text{such that}~
dist_{\srcopt\dstopt f} > \regendist. \nonumber
\end{eqnarray*}
Second, we ensure that the set of regens assigned to an IP link
indeed forms a contiguous path.  That is, for all nodes $u$ aside
from those housing the source and destination routers, the number of regens
assigned to $u$ equals the number of regens for which $u$ is the next-hop:
\begin{eqnarray*}
\sum_{v\in\nodeset} R_{\srcip\dstip uvf}
& = &
\sum_{v\in\nodeset} R_{\srcip\dstip vuf} \\
& & \forall u \in \nodeset, \forall \srcip, \dstip \in \routerset,
\forall f \in F. \nonumber
\end{eqnarray*}
We need sufficient regens at the source IP router's node \srcipnode, and sufficient
regens with the destination IP router's node \dstipnode\ as their next-hop,
for each IP link
\begin{eqnarray*}
\label{eq:regens constr first}
\sum_{\node \in \nodeset} R_{\srcip\dstip\srcipnode\node f}
& \geq & X_{\srcip\dstip f} \\
\sum_{\node \in \nodeset} R_{\srcip\dstip\node\dstipnode f}
& \geq & X_{\srcip\dstip f} \\[-0.5em]
& & \forall \srcip, \dstip \in \routerset, \forall f \in F \nonumber;
\end{eqnarray*}
But, \dstipnode\ can't have any regens, and \srcipnode\ can't
be the next-hop location for any regens
\begin{equation*}
\label{eq:no forwards backwards}
R_{\alpha\beta u\srcipnode f} =
R_{\alpha\beta\dstipnode uf} = 0
\end{equation*}
\hfill$\forall u \in \nodeset, \forall \alpha, \beta \in \routerset,
\forall f \in F.$

\para{Additional practical constraints.}
We have two practical constraints which are not fundamental to the
general problem but are artifacts of the current state of routing technology.
First, ISPs build IP links in bandwidths
that are multiples of 100 Gbps.  We encode this policy by requiring
$X_{\srcip\dstip f}$, $T_{\router}$, and $R_\node$ to be integers and converting
our demand matrix into 100 Gbps units.

Second, current IP and optical equipment require each IP link to have equal
capacity to its opposite direction.  With these constraints, only one of
\eqref{eq:t>=incoming} and \eqref{eq:t>=outgoing} is necessary.

Finally, we require all variables to take on nonnegative values.

\subsection{Dynamic Placement of IP Links}
\label{subsec:IP links}
Thus far, we have described constraints ensuring that each IP link has
enough tails and regens.  But, we have not discussed IP link placement or routing.
Although link placement and routing \emph{themselves} are part of network
operation rather than network design, they play central roles as \emph{parts} of the
network design problem.  How many are ``enough'' tails and regens for each
IP link depends on the link's capacity, and the link's capacity depends on how
much traffic it must carry.  Therefore, the network operation problem is a subproblem
of our network design optimization.

These constraints are the well-known multicommodity flow (MCF) constraints
requiring
\begin{inparaenum}[(a)]
	\item flow conservation;
	\item that all demands are sent and received; and
	\item that the traffic assigned to a particular IP link cannot exceed
	the link's capacity.
\end{inparaenum}
$Y_{st\srcip\dstip f}$ gives the amount of $(s, t)$ traffic routed
on IP link $(\srcip, \dstip)$ in failure scenario $f$.  Hence, we
express these constraints with the following equations:
\begin{align}
\label{eq:operation first}
\displaystyle\sum_{\srcip \in \routerset} Y_{stuvf} &=
\displaystyle\sum_{u \in \routerset} Y_{stvuf}
& \forall (s, t) \in D, \\[-0.8em]
&& \forall v \in \routerset - \{s, t\}, \forall f \in F \nonumber \\
\displaystyle\sum_{u \in \routerset} Y_{stsuf} &=
\displaystyle\sum_{u \in \routerset} Y_{stutf} \\
&= D_{st} & \forall s, t \in D, \forall f \in F \nonumber \\
\displaystyle\sum_{(s, t) \in D} Y_{stuvf} &\leq X_{uvf}
& \forall u, v \in \routerset, \forall f \in F. \label{eq:operation last}
\end{align}
As before, $X_{uvf}$ in Constraint \eqref{eq:operation last} is the capacity
of IP link $(u, v)$ in failure scenario $f$.

\para{Network design and operation in practice.}
Once the network has been designed, we solve the network operation problem for
whichever failure scenario represents the current state of the network by replacing
variables $T_{\router}$ and $R_{\node}$ with their assigned values.

\subsection{Extensions to a Wider Variety of Settings}
\label{subsec:extensions}
We now describe how to relax the assumptions we've made throughout
Sections \ref{subsec:objective} - \ref{subsec:IP links} that
\begin{inparaenum}[(a)]
	\item each IP node houses exactly one IP router; and
	\item traffic demands are constant.
\end{inparaenum}

\para{Accounting for multiple routers colocated at a single IP node.}
If we assume that IP links connecting routers colocated within the same IP node
always have the same cost as (short) external IP links (i.e., they require one tail at each
router endpoint), then our model already allows for any number of IP
routers at each IP node; if this assumption holds, then we simply treat colocated routers
as if they were housed in nearby nodes e.g., one mile apart.  However, in general
this assumption is not valid, because intra-IP-node links require one port per router,
rather than a full tail (combination router port and optical transponder) at each end.
Hence, intra IP node links are cheaper than even the shortest external links.
To accurately model costs we must account for them explicitly.

To do so, we add the stipulation to all the constraints presented above that, whenever
one constraint involves two IP routers, these IP routers cannot be colocated.
Then, we add the following:

Let $U$ be the set of IP routers containing $u$ and any other routers $u'$ collocated
at the same IP node with $u$.
Let $P_u$ be the number of ports placed at $u$ for intra-node links.  Let \portcost\
be the cost of one 100 Gbps port.  Our objective
function now becomes
\begin{equation*}
\min~~ \tailcost \sum_{\router \in \routerset} T_{\router} +
\regencost \sum_{\node \in \optset} R_\node +
\portcost \sum_{\router\in\routerset} P_{\router}.
\end{equation*}

Ultimately, we want to constrain the traffic traveling between $u$ and any $u'$ to fit
within the intra-node links, as follows (c.f. Constraint \eqref{eq:operation last}).
\begin{equation*}
\sum_{(s, t) \in D} Y_{stuu'f} \leq X_{uu'f}
\forall u, u' \in U, \forall U \in \ipnodeset, \forall f \in F.
\end{equation*}
 
But, no $X_{uu'f}$ appear in the objective function; the links themselves have no
defined cost.  Hence, we add constraints to limit the capacity of the links to the number
of ports $P_u$.  Specifically,
we use the analogs of \eqref{eq:t>=incoming}
and \eqref{eq:t>=outgoing} to describe the relationship between ports $P_u$ placed at
$u$ (c.f. tails placed at $u$) and the intra-node links starting from
(c.f. $X_{u\beta f}$ external IP links) and ending at (c.f. $X_{\alpha uf}$
external IP links) $u$.
\begin{eqnarray*}
\sum_{u' \in U} X_{u'uf} & \leq & P_{u} 
\label{eq:intra>=incoming} \\
\sum_{u' \in U} X_{uu'f} & \leq & P_{u}
\label{eq:intra>=outgoing} \\[-0.5em]
& & \forall U \in \ipnodeset, \forall u \in U, \forall f \in F \nonumber
\end{eqnarray*}

\para{Accounting for changing traffic.}
Thus far, we have described our model to accommodate changing failure conditions
over time with a single traffic matrix.
In reality, traffic shifts as well.  Adding this to the mathematical formulation
is trivial.  Wherever we currently consider all failure scenarios $f \in F$, we need
only consider all $(\text{failure}, \text{traffic matrix})$ pairs.  Unfortunately, while this change
is straightforward from a mathematical perspective, it is computationally costly.
The number of 
failure scenarios is a multiplicative factor on the model's complexity.  If we
extend it to consider multiple traffic matrices, the number of different
traffic matrices serves as an additional multiplier.

 \section{Scalable Approximations}
\label{sec:scalable approximations}

In theory, the network design algorithm presented above finds the optimal
solution.  We will call this approach \optimal.  However, \optimal\ does not scale,
even to networks of moderate
size ($\sim 20$ IP nodes).  To address this issue, we introduce two approximations,
\simple\ and \greedy.

\optimal\ is unscalable because, as network size
increases, not only does the problem for any given failure scenario become more
complex, but the number of failure scenarios also increases.  In a network with
$\ell$ optical spans, $n$ IP nodes, and $d$ separate demands,
the total number of variables and constraints in \optimal\ is a monotonically
increasing function
$g(\ell, n, d)$ of the size of the network and demand matrix, multiplied by the number
of failure scenarios, $\ell + n$.  Thus, increasing
network size has a multiplicative effect on \optimal's complexity.
The key to \simple\ and \greedy\
is to decouple the two factors.

\subsection{\simple\ Parallelizing of Failure Scenarios}
In \simple, we solve the placement problem separately for each
failure condition.  That is, if \optimal\ jointly considers failure scenarios
labeled
$F = \{1, 2, 3\}$, then \simple\ solves one optimization for $F = \{1\}$,
another for $F = \{2\}$, and a third for $F = \{3\}$.
The final number of tails and regens required at each site is the maximum
required over all scenarios.  Each of the $\ell + n$ optimizations is exactly
as described in Section \ref{sec:problem}; the only difference is the definition
of $F$.  Hence, each optimization has
$g(\ell, n, d)$ variables and constraints.  The problems are independent of
each other, and therefore
we can solve for all failure scenarios in parallel.  As network size
increases, we only pay for the increase in $g(\ell, n, d)$, without an
extra multiplicative penalty for an increasing number of failure scenarios.

\subsection{\greedy\ Sequencing of Failure Scenarios}
\greedy\ is similar to \simple, except we solve for
the separate failure scenarios in sequence, taking into account where tails
and regens have been placed in previous iterations.  In \simple, the $\ell + n$
optimizations are completely independent, which is ideal from a time efficiency
perspective.  However, one drawback is that \simple\ misses some opportunities
to share tails and regens across failure scenarios.  Often, the algorithm is
indifferent between placing tails at router $a$ or router $b$, so it arbitrarily
chooses one.  \simple\ might happen to choose $a$ for Failure 1
and $b$ for Failure 2, thereby producing a final solution with tails at both.  In
contrast, \greedy\ knows when solving for Failure 2 that tails have already
been placed at $a$ in the solution to Failure 1.  Thus, \greedy\ knows that a
better \emph{overall} solution is
to reuse these, rather than place additional tails at $b$.

Mathematically, \greedy\ is like \simple\ in that it requires solving $|F|$
separate optimizations, each considering one failure scenario.  But, letting
$\tap_{\router}$ represent the number of tails already placed at $\router$, we replace
Constraints \eqref{eq:t>=incoming} and \eqref{eq:t>=outgoing} with the following.
\begin{eqnarray}
\sum_{\srcip \in \routerset} X_{\srcip \router f} & \leq & T_{\router} + \tap_{\router}
\label{eq:t+tap>=incoming} \\
\sum_{\dstip \in \routerset} X_{\router\dstip f} & \leq & T_{\router} + \tap_{\router}
\label{eq:t+tap>=outgoing} \\[-0.5em]
& & \forall \node \in \routerset, \forall f \in F \nonumber
\end{eqnarray}
In \eqref{eq:t+tap>=incoming} and \eqref{eq:t+tap>=outgoing}, $T_{\router}$ represents the
number of new tails to place at router \router, not counting the $T'_{\router}$
already placed.  Similarly, with $\rap_{\node}$ defined as the number of regens already placed at
$\node$ and $R_{\node}$ as the new regens to place, Constraint \eqref{eq:r>=outgoing} becomes
\begin{equation*}
\label{eq:r+rap>=outgoing}
R_{\node} + \rap_{\node} \geq \sum_{\substack{\srcip, \dstip \in \routerset \\ \dstopt \in \optset}}
R_{\srcip\dstip\node\dstopt f} ~~ \forall \node \in \optset, \forall f \in F.
\end{equation*}

We always solve the no failure scenario first, as a baseline.  After that, we find
that the order of the remaining failure scenarios does not matter much.

With \greedy, we solve for the $\ell + n$ failure scenarios in sequence,
but each problem has only $g(\ell, n, d)$ variables and constraints.  The number of failure
scenarios is now an additive factor, rather than a multiplicative one in
\optimal\ or absent in \simple.

\subsection{Roles of \simple, \greedy, and \optimal}
\label{subsec:roles}
As we will show in Section \ref{sec:eval}, \greedy\ finds nearly equivalent-cost
solutions to \optimal\ in a fraction of the time.  \simple\ universally performs
worse than both.  We introduce \simple\ for theoretical completeness, though
due to its poor performance we don't recommend it in practice;
\simple\ and \optimal\ represent the two
extremes of the spectrum of joint optimization across failure scenarios, and
\greedy\ falls in between.

We see both \optimal\ and \greedy\ as useful and complementary tools for network
design, with each algorithm best suited to its own set of use cases.  \optimal\
helps us understand exactly how our
constraints regarding tails, regens, and demands interact and affect the
final solution.  It is best used on a scaled-down, simplified network to
\begin{inparaenum}[(a)]
\item answer questions
such as \emph{How do changes in the relative costs of tails and regens
affect the final solution?}; and
\item serve as a baseline for \greedy.
\end{inparaenum}
Without \optimal, we wouldn't
know how close \greedy\ comes to finding the optimal solution.  Hence,
we might fruitlessly continue searching for a better heuristic.
Once we demonstrate that
\optimal\ and \greedy\ find comparable solutions on topologies that both
can solve, we have confidence that \greedy\ will do a good job on
networks too large for \optimal.

In contrast, \greedy's time efficiency makes it ideally suited to place tails
and regens for the full-sized network.
In addition, \greedy\ directly models the process
of incrementally upgrading an existing network.  The foundation of \greedy\ is
to take some tails and regens as fixed and to optimize the placement
of additional equipment to meet the constraints.  When we explained \greedy, we
described these already placed tails and regens as resulting from previously considered
failure scenarios.  But, they can just as well have previously existed in the network. \section{Evaluation}
\label{sec:eval}

First, we show that CD ROADMs indeed offer savings compared
to the existing, fixed IP link technology by showing that all of
\simple, \greedy, and \optimal\ outperform current best practices
in network design.  Then we compare these three algorithms
in terms of quality of solutions and scalability.  We show that
\greedy\ achieves similar results to \optimal\ in
less time.  Finally, we show that our algorithms should
allow ISPs to meet their SLAs even during
the transient period following a failure before the network has had
time to transition to the new optimal IP link configuration.

\subsection{Experiment Setup}
\begin{figure}
\includegraphics[width=\columnwidth]{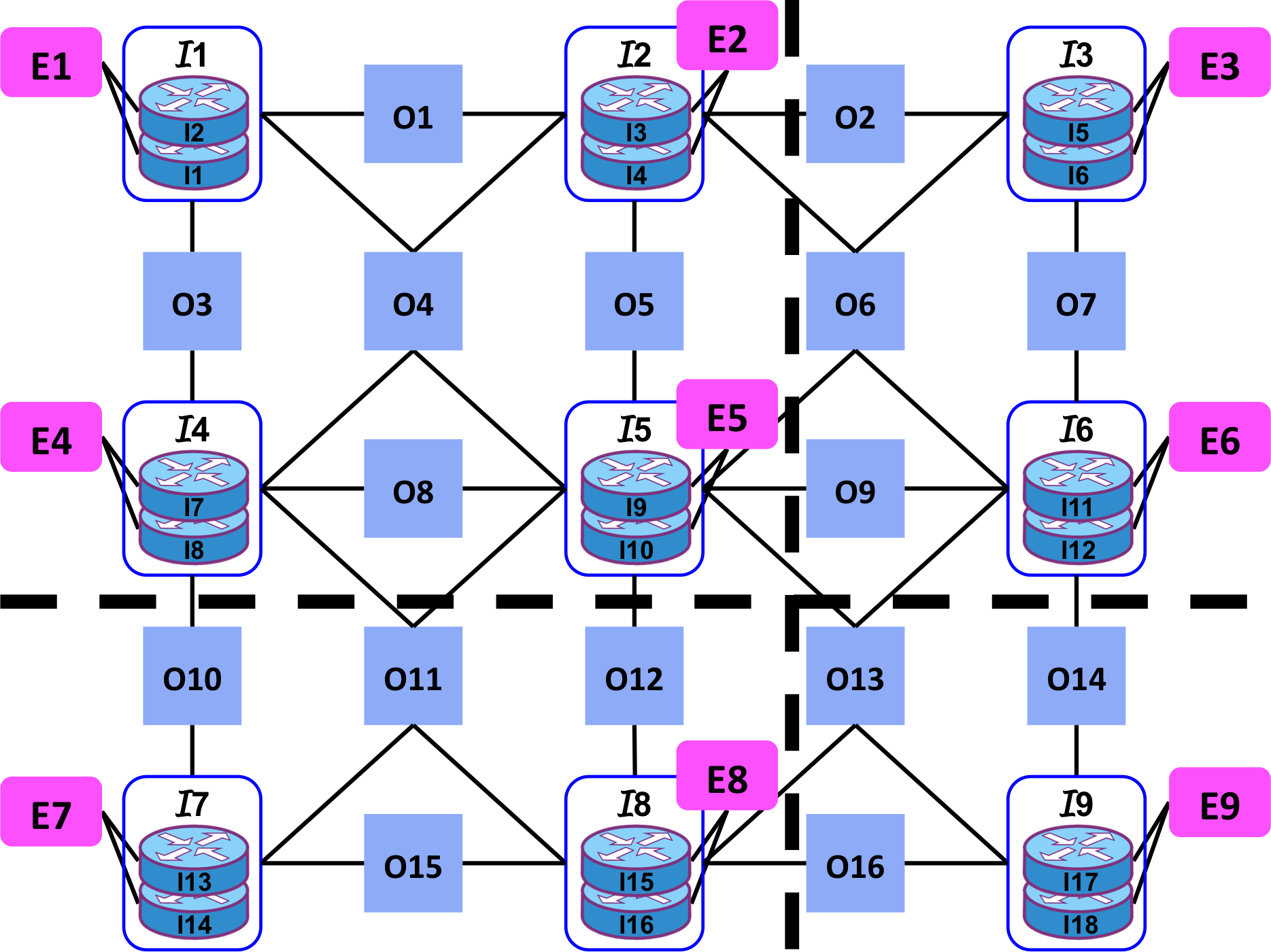}
\caption{Topology used for experiments.  We call the full network
\bigno/\bigyes, the upper two-thirds (above the thick dashed line) \medno/\medyes,
and the upper left
corner \smallno/\smallyes.}
\label{fig:Topology-9node}
\end{figure}

\para{Topology and traffic matrix.}  Figure
\ref{fig:Topology-9node}
shows the topology used for our experiments, which is representative
of the core of a backbone network of a large ISP.  The network shown in Figure
\ref{fig:Topology-9node} has nine
edge switches, which are the sources
and destinations of all traffic demands.  Each edge switch is connected to two
IP routers, which are colocated within one central office and share a single optical
connection to the outside world.
The network has an additional
16 optical-only nodes, which serve as possible
regen locations.

To isolate the benefits of our approach to minimizing tails and regens, respectively,
we create two versions of the topology in Figure \ref{fig:Topology-9node}.
The first, which we call \bigno, assigns a distance of 450 miles to each optical
span.  In this topology neighboring IP routers are only 900 miles apart, so
an IP link between them doesn't need a regen.  The second version, \bigyes,
assigns a distance of 
600 miles to each optical span.  In this topology regens are required for
any IP link.

To evaluate our optimizations on networks of various sizes, we also look at a
topology consisting of just the upper left corner of Figure \ref{fig:Topology-9node}
(above the horizontal thick dashed line and to the left of the verticle thick dashed line).
We refer to the
450 mile version of this topology as \smallno\ and the 600 mile version as \smallyes.
Second, we look at the upper two-thirds (above the thick dashed line)
with optical spans of 450 miles
(\medno) and 600 miles (\medyes).  Finally, we consider the entire topology
(\bigno\ and \bigyes).

For each topology, we use a traffic matrix in which each edge router
sends 440 GB/sec to each other edge router.  In our experiments we assume
costs of 1 unit for each tail and 1 unit for each regen, while communication
between colocated routers is free.  We use Gurobi version 8 to solve
our linear programs.

\para{Alternative strategy.}  We compare \optimal, \greedy, and
\simple\ to \legacy, the method currently used by ISPs to construct their
networks.  Once built, an IP link is fixed, and if any component fails,
the link is down and all other components previously dedicated to it
are unusable.  In our \legacy\ algorithm, we assume that IP links
follow the shortest optical path.  Similar to \greedy, we begin by
computing the optimal IP topology
for the no failure case.  We then designate those links as already paid
for and solve the first failure case under the condition that reusing any
of these links is ``free.''  We add any additional links placed in this iteration
to the already-placed collection and repeat this process for all failure scenarios.

\legacy\ is the pure IP layer optimization and failure restoration described
in Section \ref{sec:background}.  As described previously, we need not compare
our approaches to pure optical restoration, because pure optical restoration
cannot recover from IP router failures.  We need not compare against independent
optical and IP restoration, because this technique generally performs worse
than pure-IP or IP-along-disjoint-paths.

We compare against IP-along-shortest-paths, rather than IP-along-disjoint-paths,
for two reasons.  First, the main drawback
of IP-along-shortest-paths is that, in general, it does not guarantee recovery from
optical span failure.  However, on our example topologies \legacy\ \emph{can}
handle any optical failure.  Second, the formulation of the rigourous IP-along-djsoint-paths
optimization is nearly as complex as the formulation of \optimal; if we remove the restriction
that IP links must follow shortest paths, then we need constraints like those described in
Section \ref{subsec:tails regens} to place regens every 1000 miles along a link's path.  For this reason,
ISPs generally do not formulate and solve the rigorous IP-along-disjoint-paths 
optimization.  Instead, they hand-place IP links according to heuristics and historical precedent.
We don't use this approach because it is too subjective and not scientifically replicable.  In
summary, IP-along-shortest-paths strikes the appropriate balance among
\begin{inparaenum}[(a)]
	\item effectiveness at finding close to the optimal solution possible with traditional
	technology;
	\item realism;
	\item simplicity for our implementation and explanation; and
	\item simplicity for the reader's understanding and ability to replicate.
\end{inparaenum}

\begin{figure*}
\begin{subfigure}{0.5\textwidth}
  \centering
  \includegraphics[width=.8\textwidth]{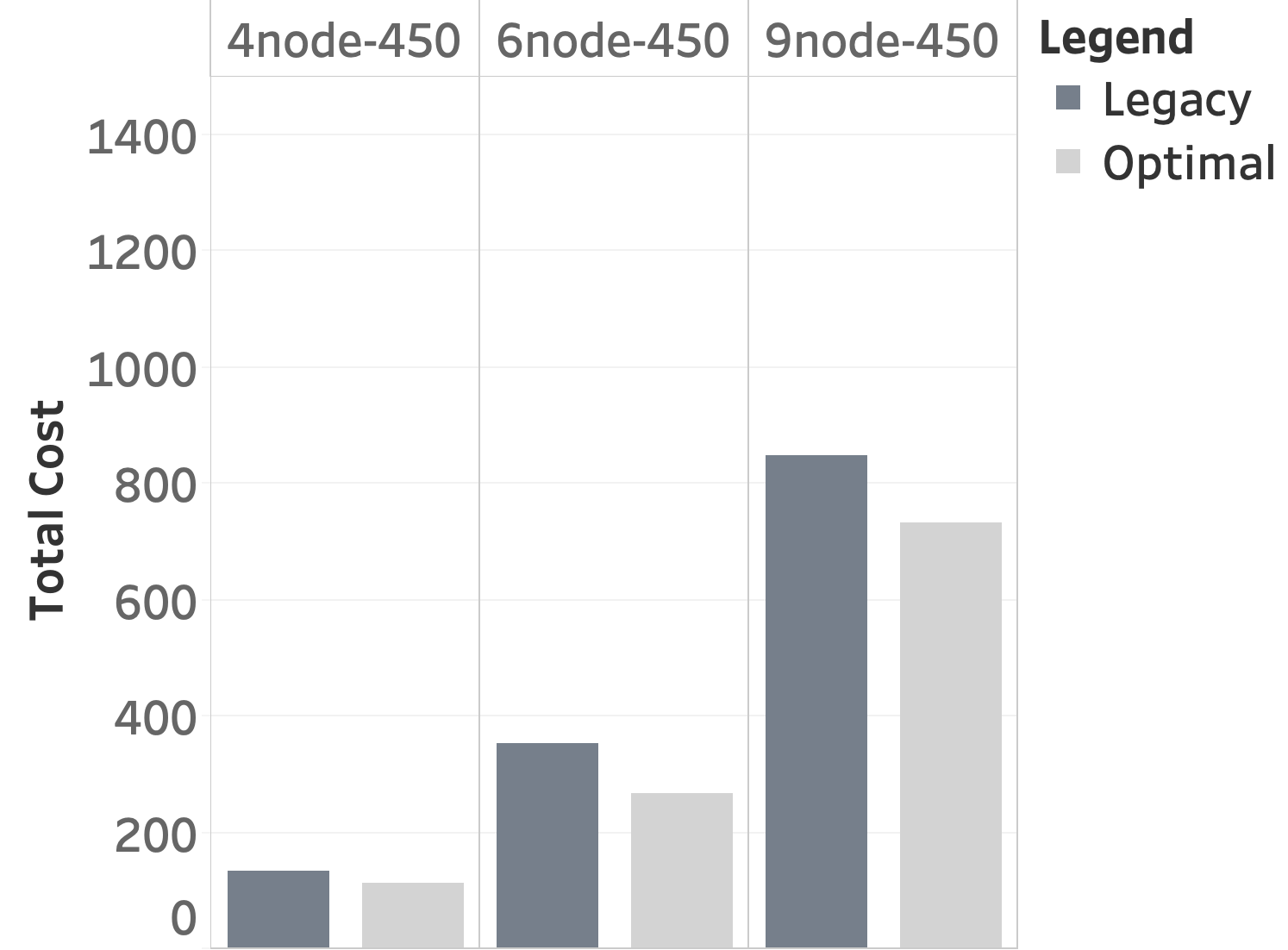}
  \caption{Neighboring optical nodes 450 miles apart.}
  \label{subfig:opt-leg-cost450}
\end{subfigure}\begin{subfigure}{.5\textwidth}
  \centering
  \includegraphics[width=0.8\textwidth]{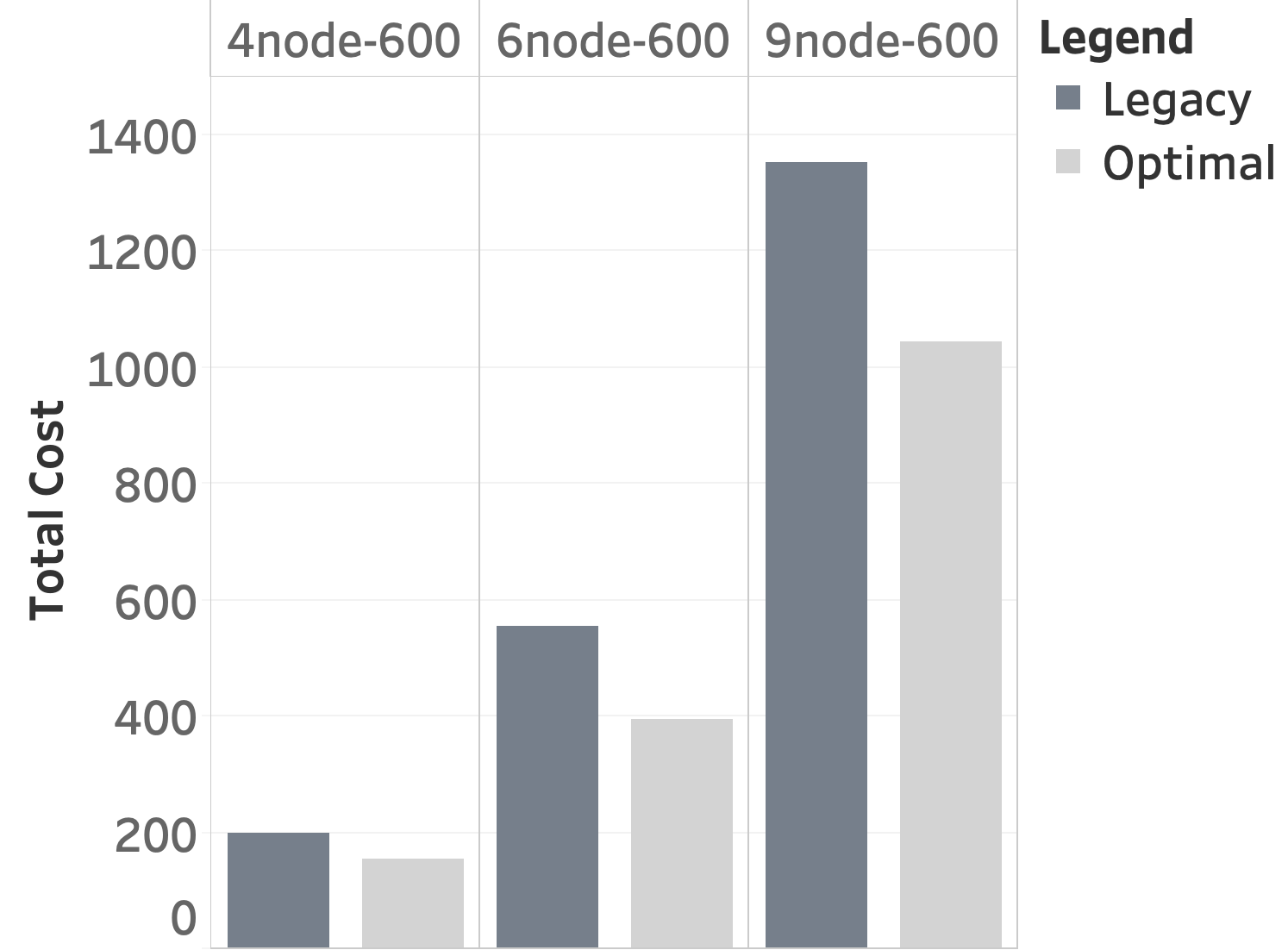}
  \caption{Neighboring optical nodes 600 miles apart.}
  \label{subfig:opt-leg-cost600}
\end{subfigure}
\caption{Total cost (tails + regens) by topology for \optimal\ and \legacy.
\optimal\ outperforms \legacy\ on all topologies, and the gap is greatest
on the largest network.}
\label{fig:cost}
\end{figure*}

\subsection{Benefits of CD ROADMs}
\label{subsec:CD ROADM benefits}

To justify the utility of CD ROADM technology, we show that building
an optimal CD ROADM network offers up to 29\% savings compared to building
a legacy network.  Since neither approach requires any regens on the 450
mile networks, all those savings come from tails.  On \smallyes\ \optimal\
requires 15\% fewer tails and 38\% fewer regens.  On \medyes\
we achieve even greater savings, using 20\% fewer tails and 44\%
fewer regens.  On \bigyes\ \optimal\ uses 16\% \emph{more} tails than
\legacy\ but more than compensates by requiring 55\% fewer regens, for an
overall savings of 23\%.  The bars in Figures \ref{fig:cost} illustrate
the differences in total cost.
Comparing Figures \ref{subfig:opt-leg-cost450}
and \ref{subfig:opt-leg-cost600}, we see that \optimal\ offers greater savings
compared to \legacy\ on the 600 mile networks.  This is because regens, moreso than
tails, present opportunities for reuse across failure scenarios.  \optimal\ capitalizes
on this opportunity while \legacy\ doesn't; both algorithms find solutions with close to
the theoretical
lower bound in tails, but \legacy\ in general is inefficient with regen placement.
Since no regens are necessary for the 450 mile topologies, this benefit of \optimal\
compared to \legacy\ only manifests itself on the 600 mile networks.

In these experiments we allow up to five minutes per failure scenario for \legacy\ and
the equivalent total time for \optimal\ (i.e., 300 sec $\times$ 21 failure scenarios =
6300 sec for \smallno\ and \smallyes, 300 sec $\times$ 35 failures = 10500 sec
for \medno\ and \medyes\ and $300 \times 59 = 17700$ sec for \bigno\ and \bigyes).

\subsection{Scalability Benefits of \greedy}
\label{subsec:greedy scalability benefits}

\begin{figure}\includegraphics[width=\columnwidth]{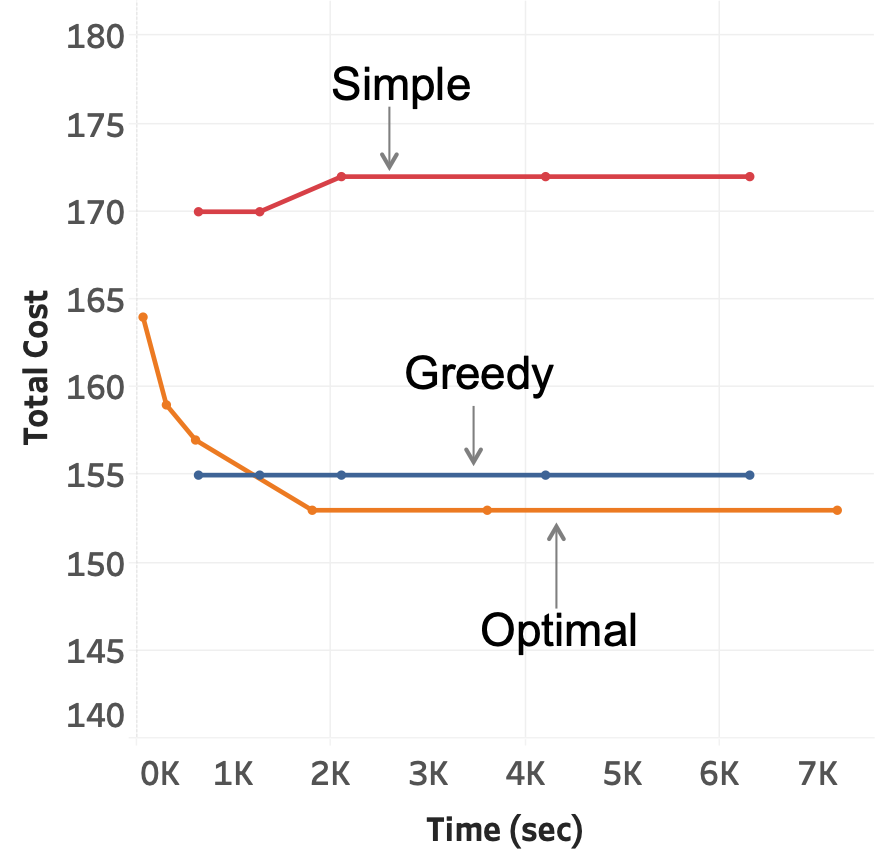}
\caption{Total cost by computation time for \simple, \greedy,
and \optimal\ on \smallyes.  Lines do not start at $t = 0$ because
Gurobi requires some amount of time to find any feasible solution.}
\label{fig:timing}
\end{figure}

As Figure \ref{fig:timing} shows, \greedy\ outperforms \optimal\
when both are limited to a short amount of time.  ``Short'' here is relative
to topology; Figure \ref{fig:timing} illustrates that the crossover point is
around 1200 seconds for \smallyes.  In contrast, both \greedy\ and
\optimal\ always outperform \simple, even at the shortest time limits.
The design \greedy\ produces costs at most 1.3\% more
than the design generated by \optimal, while \simple's design costs up to 12.4\%
more than that of \optimal\ and
11.0\% more than that of \greedy.  Reported times for these experiments do
\emph{not} parallelize \simple's failure scenarios; we show the summed
total time.  In addition, the times for \greedy\ and \simple\ are
an upper bound.  We set a time limit of $t$ seconds for each of $|F|$ failure scenario,
and we plot each algorithm's objective value at $t|F|$.

Interestingly, the objective values of \simple\, for this topology, and \greedy\ for
some others, do not monotonically
decrease with increasing time.  We suspect this is because their solutions for
failure scenario $i$ depend on their solutions to all previous failures.  Suppose that,
on failure $i - j$, Gurobi finds a solution $s$ of cost $c$ after 60 seconds.
If given 100 seconds per failure scenario, Gurobi might use the extra time
to pivot from the particular solution
$s$ to an equivalent cost solution $s'$, in an endeavor to find a configuration with
an objective value less than $c$ on this particular iteration.  Since both $s$ and $s'$
give a cost of $c$ for iteration $i - j$, Gurobi has no problem returning $s'$.  But,
it's possible that $s'$ ultimately leads to a slightly worse overall solution than $s$.
As Figure \ref{fig:timing} shows, these differences are at most 10 tails and regens,
and they occur only at the lowest time limits.

\subsection{Behavior During IP Link Reconfiguration}
\label{subsec:transient}

In the previous two subsections, we evaluate the steady-state performance of \optimal,
along with \greedy, \simple, and \legacy, after
the network has had time to transition both routing and the IP link configuration to their
new optimal settings based on the current failure scenario.  However, as we describe
in Section \ref{subsec:background CD ROADM}, there exists a period of
approximately two minutes during
which routing has already adapted to the new network conditions but IP links
have not yet finished reconfiguration.  In this section we show that our
approach gracefully
handles this transient period, as well.

The fundamental difference between these experiments and those in Sections
\ref{subsec:CD ROADM benefits} and \ref{subsec:greedy scalability benefits} is
that here we disallow IP link reconfiguration.
Whereas in Sections \ref{subsec:CD ROADM benefits} and \ref{subsec:greedy scalability benefits}
we jointly optimize both IP link configuration and routing in
response to each failure scenario, we now reoptimize only routing; for each
failure scenario we
restrict ourselves to the links that were both already established in the no-failure
case and have not been brought down by said failure.  Specifically, in these experiments
we begin
with the no-failure IP link configuration as determined by \optimal.
Then, one-by-one we consider
each failure scenario, noting the fraction of offered traffic we can carry on this topology
simply by switching from \optimal's no-failure routing to whatever is now the best
setup given the failure under consideration.

\begin{figure*}
\begin{subfigure}{0.5\textwidth}
  \centering
  \includegraphics[width=.8\textwidth]{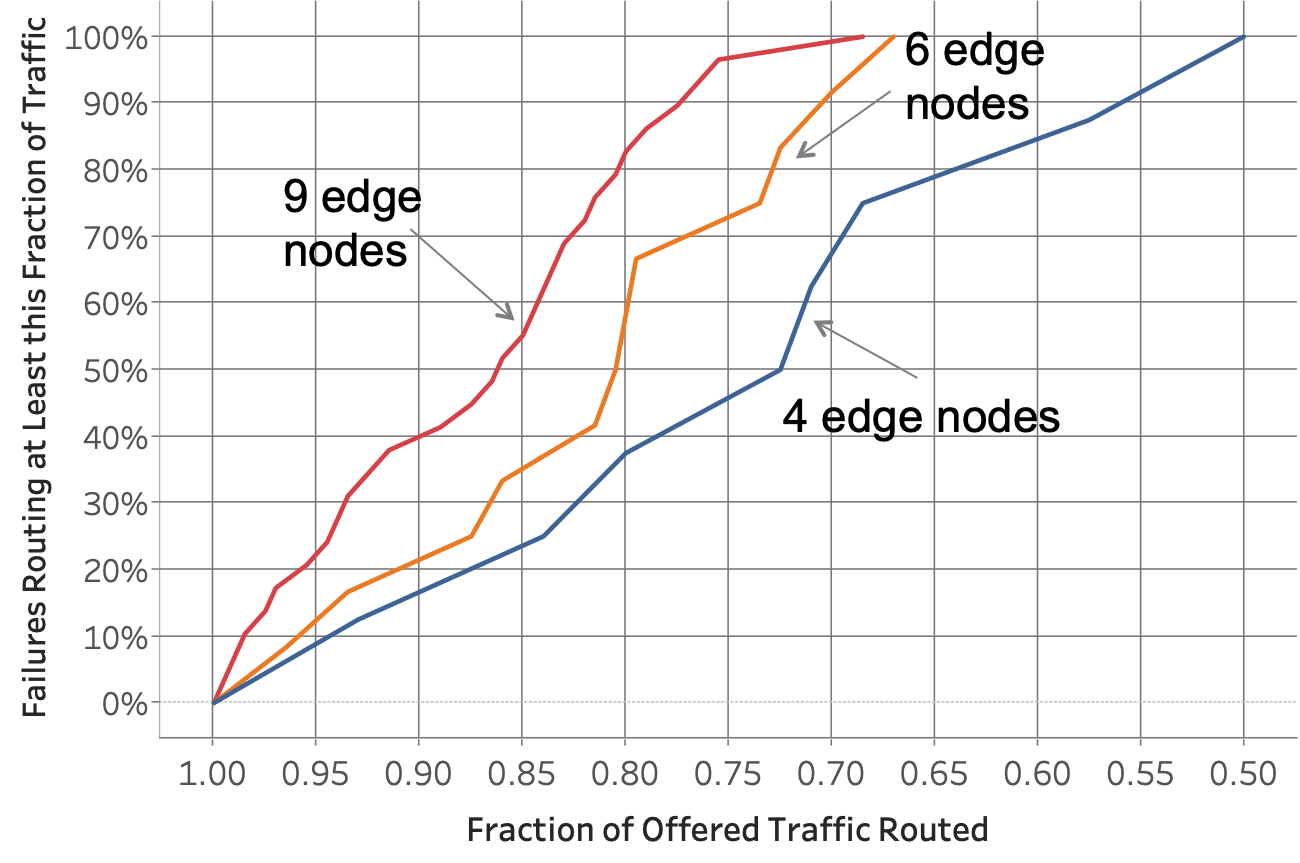}
  \caption{Neighboring optical nodes 450 miles apart.}
  \label{subfig:opt-transient450}
\end{subfigure}\begin{subfigure}{.5\textwidth}
  \centering
  \includegraphics[width=0.8\textwidth]{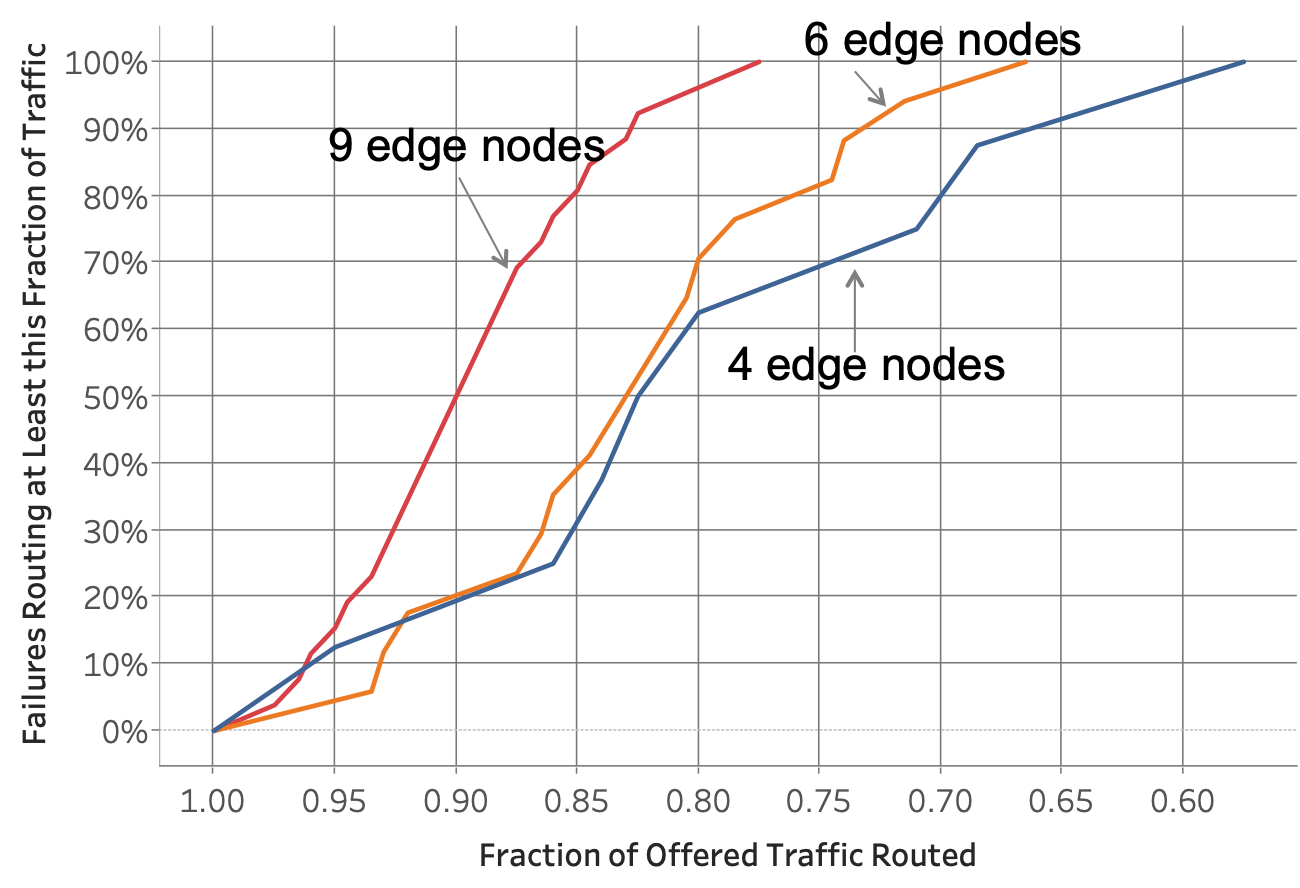}
  \caption{Neighboring optical nodes 600 miles apart.}
  \label{subfig:opt-transient600}
\end{subfigure}
\caption{Percentage of failure scenarios for which rerouting over the existing
IP links allows delivery of at least the indicated fraction
of offered traffic.}
\label{fig:transient}
\end{figure*}

Figure \ref{fig:transient} shows our results.  The graphs are CDFs illustrating the fraction
of failure scenarios indicated on the $y$-axis for which we can deliver at least
the fraction of traffic denoted by the $x$-axis.  For example, the red point
at $(0.85, 50\%)$ in Figure \ref{subfig:opt-transient450} indicates that in 50\% of the
59 failure scenarios
under consideration for \bigno, we can deliver at least 85\% of offered traffic just by
reoptimizing routing.  The blue line in Figure \ref{subfig:opt-transient450} represents
the results
of taking the 21 failure scenarios of \smallno\ in turn, and for each recording
the fraction of offered traffic routed.  The blue line in Figure \ref{subfig:opt-transient600}
shows the same
for the 21 failure scnarios of \smallyes, while the orange lines show the 35 failure
scenarios for \medno\ and \medyes, and the red lines show the 59 failure scenarios
for the large topologies.

We find two key takeaways from Figure \ref{fig:transient}.  First, across all six topologies
we always deliver at least 50\% of traffic.  Second, our results improve as the number of
nodes
in the network increases,
and we do better on the topologies requiring regens than on those that don't.  On
\bigyes, we're always able to route at least 80\% of traffic.  Generally, ISPs'
SLAs require them to always deliver all high priority traffic, which
typically represents about 40-60\% of total load.  However, in the presence of failures
or extreme congestion they're allowed to drop
low priority traffic.  Since most operational backbones
are larger even than our \bigyes\ topology, our results suggest that our algorithms
should always allow ISPs to meet their SLAs.  Note that we don't expect to be
able to route 100\% of offered traffic in all failure scenarios without reconfiguring IP
links; if we could there would be little reason to go through the reconfiguration process at
all.  But, we already saw in Section \ref{subsec:CD ROADM benefits} that remapping the
IP topology to the optical
underlay adds significant value.

\section{Related Work}
\label{sec:related}

Though there has been significant work on various aspects of IP/optical
networks, no existing research addresses the joint optimization of IP
and optical network design.

At a high level, the Owan work by Jin et al. \cite{Jin2016} is similar to ours.
Like our work, Owan is a centralized system that jointly optimizes the IP and optical
topologies and configures network devices, including CD ROADMs, according to this
global strategy.  However,
there are three key differences between Owan and our work.
 
 First, our objective differs from that of Jin et al.  We aim to minimize the cost of
 tails and regens such
that we place the equipment such that, under all failure scenarios,
we can set up IP links to carry all necessary traffic.  Jin et al. aim to
minimize the transfer completion time or maximize the number
of transfers that meet their deadlines.
 
 Second, our work applies in a different setting.  Owan is designed for bulk transfers
 and depends on the network operator being able to control sending rates,
 possibly delaying traffic for several hours.
 We target all ISP traffic;
we can't rate control any traffic, and we must route all demands, even in the case
of failures, except during a brief transient period during IP link reconfiguration.

Third, we make different assumptions about what parts of the infrastructure are
given and fixed.  Jin et al. take the locations of optical equipment as an input
constraint, while we solve for the optimal places to put tails and regens.  This
distinction is crucial; Jin et al. don't need any notion of here-and-now decisions about
where to place tails and regens separate from wait-and-see
decisions about IP link configuration and routing.
 
Other studies
demonstrate that, to minimize delay, it is best to set up direct IP links between
endpoints exchanging significant amounts of traffic, while relying on packet
switching through multiple hops to handle lower demands \cite{Brzezinski2005}.

Choudhury \cite{Choudhury2018} and Jin \cite{Jin2016} consider joint IP/optical
optimization but use heuristic algorithms.

\section{Conclusion}
\label{sec:conclusion}
Advances in optical technology and SDN have decoupled IP links from their underlying
infrastructure (tails and regens).  We have precisely stated and solved the new network
design problem deriving from these advances, and we have also presented a fast
approximation algorithm that comes very close to the optimal solution.

\section*{Acknowledgment}
The authors would like to thank Mina Tahmasbi Arashloo for her discussions
about the regen constraints and Manya Ghobadi, Xin Jin, and Sanjay Rao
for their feedback on drafts.

\bibliographystyle{IEEEtran}
\bibliography{bibliography,supplemental}

\iffalse

\begin{IEEEbiography}{Michael Shell}
Biography text here.
\end{IEEEbiography}

\begin{IEEEbiographynophoto}{John Doe}
Biography text here.
\end{IEEEbiographynophoto}

\begin{IEEEbiographynophoto}{Jane Doe}
Biography text here.
\end{IEEEbiographynophoto}

\fi
\end{document}